\documentclass[eqsecnum,aps,pra,twocolumn,groupedaddress,showpacs,showkeys]{revtex4}

\usepackage{graphicx}
\begin{document}

\newcommand{\be}{\begin{equation}}
\newcommand{\ee}{\end{equation}}

\title{Relativistic mechanics of Casimir apparatuses in a weak
gravitational field}

\author{Giuseppe Bimonte, Enrico Calloni, Giampiero Esposito and Luigi Rosa}

\affiliation{Dipartimento di Scienze Fisiche, Universit\`{a} di
Napoli Federico II, Complesso Universitario MSA, Via Cintia
I-80126 Napoli, Italy;\\ INFN, Sezione di Napoli, Napoli, ITALY\\
}

\date{\today}

\begin{abstract}

This paper derives a set of general relativistic Cardinal Equations for the
equilibrium of an extended body in a uniform gravitational field.
These equations are essential for a proper understanding of the
mechanics of suspended relativistic systems. As an example, the
prototypical case of a suspended vessel filled with radiation is discussed.
The mechanics of Casimir apparatuses at rest in the gravitational field of the
Earth is then considered. Starting from an
expression for the Casimir energy-momentum tensor
in a weak gravitational field recently derived by the authors,
it is here shown that, in the case of a
rigid cavity supported by a stiff mount, the weight of the Casimir
energy $E_C$ stored in the cavity corresponds to a gravitational
mass $M=E_C/c^2$, in agreement with the covariant conservation law of
the regularized energy-momentum tensor.
The case of a cavity consisting of two
disconnected plates supported by separate mounts, where the two
measured forces cannot be obtained by straightforward arguments,
is also discussed.

\end{abstract}


\maketitle

\section{Introduction}

One of the most intriguing predictions of Quantum Electrodynamics
is the existence of irreducible fluctuations of the
electromagnetic (e.m.) field  in the vacuum. It was Casimir's
fundamental discovery \cite{casimir} to realize that the effects
of this purely quantum phenomenon were not confined to the atomic
scale, but would rather manifest themselves also at the
macroscopic scale, in the form of an attractive force between two
parallel discharged metal plates at distance $a$. Under the
simplifying assumption of perfectly reflecting mirrors, he
obtained a force of magnitude \be F_{\rm (C)}=\frac{\pi^2}{240}
\frac{\hbar \; c}{a^{4}} {\cal A}, \label{paral} \ee where ${\cal
A}$ is the area of the plates. By modern experimental techniques
the Casimir force has now been measured with an accuracy of a few
percent (see Refs. \cite{decca} and Refs. therein). For detailed
reviews of both theoretical and experimental aspects of the
Casimir effect, see Refs.
\cite{bordag,Milt04,Nest04,Lamo05,Capa07}.

Now, the energy  associated with the Casimir force in Eq.
(\ref{paral}) is
\be
E_{\rm (C)}=-\frac{\pi^2}{720} \frac{\hbar \; c}{a^{3}}{\cal A},
\label{ener}
\ee
and one may wonder if it is
possible to measure this vacuum energy directly, rather than the
corresponding force. Experiments of this sort would further
enhance one's confidence in the reality of vacuum fluctuations.
Recently, we have proposed an experiment with superconducting cavities,
aiming at measuring the {\it variation} of Casimir energy that
accompanies the superconducting transition \cite{alad}. Another
line of research concerns the gravitational coupling of the vacuum
energy. This problem has been studied by a number of authors,
leading to some contradictory conclusions.

In Ref. \cite{Kari00}, the
authors propose an experiment to measure the Casimir force in the
Schwarzschild metric of the galactic center. The experiment is
designed to show whether or not virtual quanta follow geodesics.
They find gravitational forces that depend on the orientation of
the Casimir apparatus with respect to the gravitational field of
the earth.

In Ref. \cite{Cald02}, the author evaluates scalar Casimir effects
in a weak gravitational field, and obtains corrections to the
vacuum energy-momentum tensor and attractive force on the plates,
resulting from spacetime curvature. He then points out that, if
the cosmological constant arises by virtue of zero-point energy,
it is susceptible to fluctuations induced by gravitational
sources. He uses a curved line element which describes the weak
gravitational field in the vicinity of a mass $M$ as a
perturbation to Minkowski spacetime rather than the  flat  metric
appropriate for a uniform gravitational field, in the Fermi
coordinate system attached to the cavity.

In Ref. \cite{Sorg05}, the author studies the Casimir
vacuum energy density for a massless scalar field confined between
two nearby parallel plates in a slightly curved, static spacetime
background, employing the weak-field approximation, and obtains
the gravity-induced correction to Casimir energy. He then finds
that the attractive force between the cavity walls is expected to
weaken.

In Ref. \cite{cal}, a sketchy computation is presented to show
that, if one suspends a {\it rigid} Casimir cavity, the vacuum
fluctuations contribute an extra {\it negative} weight $\vec{
P}^{\rm(C)}$ which, to leading order in the dimensionless
parameter $\epsilon \equiv 2 {\rm g}\, a/c^2$, is equal to \be
\vec{P}^{\rm(C)}= \,\frac{ E_{(\rm C)}}{c^2}\,\vec{{\rm g}},
\label{pesoc} \ee where ${\vec {\rm g}}$ is the gravity
acceleration. In the same paper, the feasibility of such an
experiment is also discussed. An important progress was made in
Ref. \cite{bimonte}, which contains the first detailed
quantum-field-theoretic computation of the Casimir energy-momentum
tensor $\langle T^{ab}_{\rm(C)}\rangle$ (with angle brackets
denoting the vacuum expectation value) in a weak gravitational
field. This calculation appears rather important since, for
quantum field theory in curved spacetime \cite{Full89}, a
fundamental task, if not the main problem, is to understand the
energy-momentum tensor (see the Introduction in Ref.
\cite{DeWi75}). To check consistency, covariant conservation of
$\langle T^{ab}_{\rm(C)}\rangle$, i.e. \be \nabla_a \langle
T^{ab}_{\rm(C)}\rangle=0, \label{equivac} \ee was explicitly
verified therein up to first order in $\epsilon$. Unfortunately,
we used incorrectly the expression for $\langle
T^{ab}_{\rm(C)}\rangle$ derived in Ref. \cite{bimonte}, Eqs.
(\ref{T00}-\ref{T33}) below, to predict a net push on the Casimir
apparatus that is {\it four} times the correct value, given in Eq.
(\ref{pesoc}), in contradiction with what was stated in Ref.
\cite{cal}. This discrepancy was stressed in a recent paper Ref.
\cite{Full07}, where valuable variational methods are used to show
that, for the case of parallel conducting plates, the Casimir
energy gravitates according to Eq. (\ref{pesoc}). This is in
agreement with the early findings of Jaekel and Reynaud, who
studied the inertia of Casimir energy in two dimensions
\cite{Jaek93}. No orientation dependence has been found in Ref.
\cite{Full07}, accounting clearly for the discrepancy in this
respect as compared with the findings in Ref. \cite{Kari00}. The
work in Ref. \cite{Full07} and the lack of careful force formulae
in our paper \cite{bimonte}, led us to undertake a fresh thorough
mechanical analysis of the relativistic mechanics of Casimir
apparatuses at rest in the gravitational field of the Earth,
presented in the next Sections. The problem turns out to be rather
subtle, and the most significant result of a careful analysis is
that the energy-momentum tensor derived in \cite{bimonte} does
indeed lead to the correct result for the gravitational push, as
given in Eq. (\ref{pesoc}), in agreement with the findings of Ref.
\cite{Full07}.

Our paper reinterprets therefore the work in Refs. \cite{cal, bimonte}.
The content of the paper can be divided into two main parts. The
first part, coinciding with Secs. II and III, discusses the
mechanics of an extended body at rest in a uniform gravitational
field, within Einstein's Theory of General Relativity. The
analysis starts from the assumption that the body satisfies the
covariant conservation law expressing the balance of energy and
momentum between the body and the fields, Eq. (\ref{eom}) below,
and we use this to obtain a simple mathematical proof of the
general global conditions, the Cardinal Equations, that ensure
mechanical equilibrium of the body. Conditions similar to ours
were originally obtained in Refs. \cite{nord, fearn}, in the
context of any theory of gravitation satisfying the weak
Equivalence Principle, by means of an ingenious {\it gedanken}
experiment. Using these Cardinal Equations, we show that {\it any}
system, which obeys  Eq. (\ref{eom}) below, possesses a passive
gravitational mass that is equal to its total inertia. In Sec. III
we illustrate the conditions obtained in Sec. II to study the
mechanical forces in the prototypical case of a rigid suspended
vessel, filled with a fluid. The general conditions derived in
Sec. II are indispensable for a correct understanding of the
forces in a relativistic system, like a vessel filled with
radiation, considered in Sec. III, and even more so in the case of
Casimir apparatuses, that are studied in  Sec. IV.

In the second part of the paper, coinciding with Sec. IV, we study
the relativistic mechanics of Casimir apparatuses at rest in the
gravitational field of the Earth. Since from our Ref.
\cite{bimonte}, we now know that vacuum fluctuations satisfy
Eq. (\ref{equivac}), the
general theorems derived in the first two sections can be used. In
particular, we evaluate the forces exerted by the mounts that hold
the apparatus, which represent the actual quantities to be
measured in a real experiment. The problem turns out to be rather
subtle, because, according to the general Cardinal Equations, the
magnitudes of the supporting forces depend on {\it where} they are
applied, and therefore the answer depends on the setting
considered for the mounts. This is a natural phenomenon in general
relativity, as it was
discovered long ago by Nordtvedt \cite{nord}, and it turns out to
be of fundamental importance in the analysis of an essentially
relativistic system like a Casimir cavity. We consider two
different settings for the mounts. In the first case, we have just
one mount, supporting  a {\it rigid} cavity; for this case, the
general theorems in Sec. II  immediately lead to Eq. (\ref{pesoc})
for the ``weight'' of the Casimir apparatus. In the second setup,
the plates of the Casimir cavity are {\it disconnected}, and they
are supported by separate mounts. Using  the expression for
$T^{ab}_{\rm(C)}$ computed in Ref. \cite{bimonte}, we obtain the
forces exerted by the plates of the respective mounts. Sec. V
finally contains our conclusions and a discussion of the results.

\section{Relativistic Static Cardinal Equations in a uniform
gravitational field}

In this Section, we obtain a rigorous proof,
within the context of Einstein's
General Theory of Relativity, of the conditions for mechanical
equilibrium of an extended body at rest in a {\it uniform}
gravitational field.

In General Relativity, the equations expressing
the balance of energy and momentum between a body and the fields are
\be
\nabla_a T^{ab} = f^b_{(\rm vol)},
\label{eom}
\ee
where $T^{ab}$ is the energy-momentum tensor and $f^a_{(\rm vol)}$
are the external forces.
If the system is also subject to forces $f^a_{(\rm sur)}$,
applied at points of its surface $\partial \Sigma$,  Eqs.
(\ref{eom}) are supplemented by the following boundary conditions:
\be
T^{ab} n_b= -f^a_{(\rm sur)} \; {\rm on}\;\;
\partial \Sigma ,
\label{bc}
\ee
where $n^a$ is the unit normal at
$\partial \Sigma$ pointing outwards the body.

Now we consider the case of a body at rest in a uniform
gravitational field. As we shall see, in this special case it is
possible to derive, from the local conditions Eqs. (\ref{eom}) and
Eq. (\ref{bc}), a set of {\it global} conditions ensuring
mechanical equilibrium of the body. The global conditions that we
shall obtain have a form similar to the familiar Cardinal
Equations of Statics in Newtonian Theory, and will provide us with
a relativistic concept of weight for an extended body.
A striking point of departure from
classical theory is however the fact, first discovered in Ref.
(\cite{nord}), that the weight of a body, intended as the
magnitude of the force that must be applied to hold it, depends on
where the force is applied.

In what follows, Latin letters from the beginning of the alphabet
will be used as abstract indices for tensors, while Greek letters
and Latin letters from the middle of the alphabet will denote
spacetime and space components of tensors in a
definite coordinate system, respectively.

We begin by defining a uniform gravitational field as the field
seen in a uniformly accelerating frame. As is well known
\cite{MTW} in Einstein's General Theory of Relativity, such a
field is described by the line element
\be
ds^2=
-c^2\left(1+\frac{A \,z}{c^2}\right)^2 \,dt^2+\delta_{ij} \,dx^i d
x^{j} ,
\label{un}
\ee
with $A>0$ the acceleration parameter. This
line element  describes also the gravitational field of the Earth,
in a local Fermi coordinate system, once tidal effects are
neglected.

We denote by $u^a$ the normalized velocity field for observers at rest
in the metric (\ref{un}):
\be
u^a=u^0
\left(\frac{\partial}{\partial t}\right)^a=
\frac{c}{|g_{00}|^{1/2}}\left(\frac{\partial}{\partial t}\right)^a .
\label{u0}
\ee
They possess an acceleration $a^a$ in
the upwards $z$-direction, i.e.
\be
a^a=\frac{D u^a}{D
\tau}=\frac{c^2}{|g_{00}|^{1/2}}\,\partial_z\,|g_{00}|^{1/2}
\hat{z}^a=\frac{A}{1+A \,z/c^2}\, \hat{z}^a .
\label{acc}
\ee

We define the gravity acceleration ${\rm g}^a$ as {\it minus} the
acceleration $a^a$ of stationary observers, i.e.
\be
{\rm g}^a (z)
\equiv -\frac{D u^a}{D \tau}= -\frac{A}{1+A \,z/c^2}\,
\hat{z}^a .
\label{g}
\ee

We consider an extended body at rest in the coordinate system of
Eq. (\ref{un}). The body being at rest, one has
\be
T^{ab}= \rho
\, u^a u^b + S^{ab},
\label{tab}
\ee
where $\rho$ is the
{\it inertial mass} density and $S^{ab}$ is the stress tensor,
satisfying the condition
\be
S^{a}_b\; u^b=0 .\label{su}
\ee
By virtue of Eq. (\ref{u0}), we then have
\be
S^{00}=S^{0i}=S^{i0}=0.
\label{(2.9)}
\ee
and therefore $S^{ab}$ is purely spatial.

If we now insert Eq. (\ref{tab}) into the l.h.s. of Eq. (\ref{eom}),
we obtain
\be
\nabla_a (\rho \, u^a)\,u^b- \rho \,{\rm
g}^b=-\nabla_a S^{ab}+ f^b_{(\rm vol)}.
\label{eom2}
\ee
Let now $e_{(i)}^a$ be the vector fields for the coordinate spatial axis:
\be
e_{(i)}^a=\left(\frac{\partial}{\partial
x^i}\right)^a\;\;\;i=x,y,z .
\label{(2.11)}
\ee
Upon multiplying Eq. (\ref{eom2}) by $e_{(i)b}$, we obtain
\be
-\rho \, {\rm g}_i=
-\nabla_a(S^{ab}e_{(i)b})+ S^{ab}\nabla_{(a}e_{(i)b)} +f_{(\rm
vol) i}\;,
\label{eom3}
\ee
where we define ${\rm g}_i \equiv {\rm
g}_a\,e_{(i)}^a$. A simple computation gives
\be
\nabla_{(a}e_{(i)b)}=-\delta_{i\, 3}\, A \,(d t)_a \,(d t)_b ,
\label{(2.13)}
\ee
and since $S^{ab} (d t)_b=0$ (see Eq. (\ref{su})), we see that the
second term on the r.h.s. of Eq. (\ref{eom3}) vanishes. On the
other hand, for the first term on the r.h.s. of Eq. (\ref{eom3}),
using well known identities, we find
\begin{eqnarray}
\; & \; &
\nabla_a(S^{ab}e_{(i)b})=\frac{1}{\sqrt{|g|}}\,\partial_j \,
(\sqrt{|g|}\, S^{j}_i) \nonumber \\
&=& \frac{1}{\sqrt{|g_{00}|}}\,
{\partial}_j \,(\sqrt{|g_{00}|}\, S^j_i).
\label{stress}
\end{eqnarray}
Therefore, Eq. (\ref{eom3}) becomes
\be
-\rho \, {\rm g}_i= -
\frac{1}{\sqrt{|g_{00}|}}\, {\partial}_j \,(\sqrt{|g_{00}|}\,
S^j_i)+f_{(\rm vol) i}.
\label{eom3bis}
\ee
Let us now introduce
the gravitational red-shift $r_{{\small O}}(P)$  of the point $P$
with coordinates $\{x,y,z\}$ relative to an arbitrary point $O$
with coordinates $\{x_{(O)},y_{(O)},z_{(O)}\}$:
\be
r_{O}(P)=\sqrt{\frac{|g_{00}(P)|}{|g_{00}(O)|}}=\frac{1+A\, z
/c^2}{1+A\, z_{(O)} /c^2} .
\label{reds}
\ee
Upon multiplying Eq. (\ref{eom3bis}) by $r_{O}(z)$ we obtain
\be
   -\rho \,  r_{O}\, {\rm g}_i(z) =  -
 {\partial}_j ( r_{O} {S^{j}}_i) +
 r_{O} f_{(\rm vol) i}.
\label{eom4}
\ee
However, in view of Eq. (\ref{g}), se see that
\be
r_{O}(z)
\,{\rm g}_i(z)= {\rm g}_i(z_{(O)}),
\label{gzero}
\ee
and hence we arrive at the following equation:
\be
-\rho \,{\rm g}_i(z_{O})=-
{\partial}_j ( r_{O} {S^{j}}_i)  + r_{O}\, f_{(\rm vol)
i}.
\label{eom5}
\ee
Upon integrating the above equation over the
body's volume, and in view of Eqs. (\ref{bc}), we obtain our First
Cardinal Equation:
\be
\vec{P}_O+ \vec{F}_{O}^{(\rm
tot)}=\vec{0}.
\label{card1}
\ee
In this equation, the total external force is
\be
{\vec F}_{O}^{(\rm
tot)}\equiv \int_{\Sigma} d^3 x \,r_{O} \vec{f}_{(\rm vol)} +
\int_{\partial \Sigma} d^2 \sigma \, r_{O} \vec{f}_{(\rm sur)},
\label{ftot}
\ee
while the weight $\vec{P}_O$ is defined as
\be
\vec{P}_O \equiv M\, \vec{{\rm g}}_O ,
\label{wei}
\ee
where $\vec{{\rm g}}_O$ denotes the gravity acceleration at $O$ and
\be
M= \int_{\Sigma} d^3 x\, \rho .
\label{mass}
\ee
Now the quantity
$M$ in Eq. (\ref{wei}) is, by definition, the passive
gravitational mass of the body, and therefore Eq. (\ref{mass})
tells us that $M$ is equal to the total inertia of the body. We stress that
the above derivation shows that this identity is a necessary consequence of
the covariant equations (\ref{eom}).
Therefore, {\it for all physical systems which
satisfy  Eq. (\ref{eom}), the passive gravitational mass is equal
to the total inertia.}

In order to derive the Second Cardinal Equation, we now define the
center of mass $\vec{x}_{CM}$ via the equation
\be
\vec{x}_{CM}=\frac{1}{M}\, \int_{\Sigma} d^3 x \, \rho\,
\vec{x}.
\label{com}
\ee
Now we multiply Eq. (\ref{eom5}), for
$O=CM$, by $\epsilon_{kli}(x-x_{CM})^l$ and integrate the
resulting Equation over the body's volume. On using the identity
\be
\epsilon_{kli}(x-x_{CM})^l\,{\partial}_j ( r_{CM}
{S^{j}}_i)={\partial}_j[\epsilon_{kli}(x-x_{CM})^l\,  r_{CM}
{S^{j}}_i],
\label{(2.25)}
\ee
implied by the symmetry of $S^{ij}$, one obtains
the Second Cardinal Equation
\be
\vec{\tau}_{CM}^{(\rm
tot)}=\vec{0},
\label{card2}
\ee
where  $\vec{\tau}_{CM}^{(\rm
tot)}$ is the total torque of the external forces, relative to
to the center of mass:
\begin{eqnarray} &&\vec{\tau}_{CM}^{(\rm
tot)}=\int_{\Sigma} \!\!\!d^3 x \, (\vec{x}-\vec{x}_{CM}) \times
r_{CM}\,\vec{f}_{(\rm vol)} + \nonumber \\
&&+\int_{\partial \Sigma}\!\!\!\!\! d^2 \sigma
\,(\vec{x}-\vec{x}_{CM}) \times r_{CM}\,\vec{f}_{(\rm sur)}.
\label{tautot}
\end{eqnarray}
Equations similar to Eqs.
(\ref{card1}-\ref{ftot}) were first obtained in Ref. \cite{nord},
by exploiting the phenomenon of
gravitational red-shift for photons, via an ingenious {\it
gedanken} experiment involving an ideal electro-mechanical device
converting into photons the mechanical work done by a heavy body,
as it lowers or rises into the gravitational field. By a similar
procedure, the authors of Ref. \cite{fearn} obtained equations
similar to our Eqs. (\ref{card2}-\ref{tautot}), ensuring
rotational equilibrium of the body.

As we see, Eqs. (\ref{card1}-\ref{tautot}) look remarkably similar
to the analogous Equations of Newtonian theory. The striking
difference with respect to classical theory is that {\it when forces and
torques are added, one has to multiply each of them by the
red-shift of the point where they act, relative to the point where
they are added}. A remarkable consequence of this is that the
force that must be applied to a body to hold it still in a
gravitational field, depends on {\it where} the force is applied
\cite{nord}. To see this, define the {\it proper} weight
${\vec P}$ of a body as
\be
{\vec P} \equiv M\, \vec{{\rm g}}_{CM},
\label{(2.28)}
\ee
and suppose that the supporting force ${\vec f}_Q$ is applied at the
point $Q$. Then, from Eq. (\ref{card1}) we obtain
\be
{\vec f}_Q=-M\, \vec{{\rm g}}_{Q}=-r_{Q}(CM)\,{\vec P},
\label{(2.29)}
\ee
where in the final passage we used Eq. (\ref{gzero}), for $O=Q$ and $P=CM$.
Therefore, the magnitude of the applied force is equal to the
proper weight, multiplied by the red-shift of  the center of mass
$CM$ relative to the point $Q$.

We would like to comment now on an alternative possible form of
the Cardinal Equations, based on a different rearrangement of Eq.
(\ref{eom3bis}). As we shall see, the alternative form implies a
definition of inertia for a stressed body, which explicitly
involves the stresses. For this purpose, we have to consider again
the last line of Eq. (\ref{stress}). If we perform the partial
derivatives, and use Eqs. (\ref{acc}) and (\ref{g}), we obtain
\be
\frac{1}{\sqrt{|g_{00}|}}\, {\partial}_j \,(\sqrt{|g_{00}|}\,
S^j_i)= {\partial}_j \, S^j_i - \frac{1}{c^2}\, S^{j}_i\,{\rm
g}_j(z).
\label{parder}
\ee
If we substitute this expression into
Eq. (\ref{eom3bis}), and rearrange the terms, we obtain
\be
-\left(\rho\,\delta_i^j+\frac{1}{c^2}\,S^j_i \right)\, {\rm g}_j=
-   {\partial}_j \,  S^j_i+f_{(\rm vol) i}.
\label{eomstr}
\ee
This form of the equation suggests that the inertia of a stressed
body is not just $\rho$, but rather is a tensor $m^i_j$ \cite{MTW}
that depends on the stresses, i.e.
\be
m^i_j=\rho\,\delta^i_j+\frac{1}{c^2}\,S^i_j .
\label{mij}
\ee
If we integrate this equation over the body's volume, instead of Eq.
(\ref{card1}), we obtain
\be
\vec {\cal P}+\vec{\cal F}^{({\rm tot})}=0,
\label{card1bis}
\ee
where the ``weight'' $\vec{\cal P}$
is now defined as
\be
{\vec {\cal P}} \equiv \int_{\Sigma} d^3 x\;
m^j_i \, {\rm g}_j(z)\,\hat{x}^i ,
\label{we}
\ee
and $\vec{\cal F}^{({\rm tot})}$
coincides with the classical expression for the
total external force:
\be
\vec{\cal F}^{({\rm tot})}=\int_{\Sigma}
d^3 x \, \vec{f}_{(\rm vol)} + \int_{\partial \Sigma} d^2 \sigma
\, \vec{f}_{(\rm sur)} .
\label{ftotcl}
\ee
It is of course
possible to derive from Eq. (\ref{eomstr}), by similar steps as
those followed earlier, the analogue of the Second Cardinal
Equation (\ref{card2}), but we shall not write it here. We would
like to comment, instead, on the conceptual differences between
the two approaches. Indeed, the key difference arises from the
fact that the two approaches use different expressions for the sum
of forces acting at different points and, in particular, for the
sum of the contact forces acting on the faces of an infinitesimal
cube inside the body. Consider a small cube with vertex at
$\{x,y,z\}$ and sides $\{dx,dy,dz\}$, and consider a pair of
opposite faces, say the $yz$ faces at $x$ and $x+d x$. Then, by
definition of stress tensor, the forces in the direction $i$
acting on these two faces are, respectively,
$dy \,dz\,S^{x i}({x})$ and
$-dy \,dz\,S^{x i}({x}+d x)$. The question now is:
what do we take for the sum of these two forces? If we thrust the
picture outlined in Ref. \cite{nord} according to which the
phenomenon of red-shift for forces is a {\it physical} one, we
should say that the sum at $O$ of the elementary forces on the
$yz$ faces is
\begin{eqnarray}
\; & \; & -dy \,dz\,( r_O({x}+d x) S^{x
i}({x}+dx)+r_O({x})
S^{x i}({x})) \nonumber \\
&=& -d x\,dy\,dz\,\partial_x \,(r_O({x}) S^{x i}({x}))) .
\label{(2.36)}
\end{eqnarray}
Then, upon summing over the three pairs of faces, we obtain
for the total force $d F^i_O(x)$ the expression
\be
dF^i_O=-
dx\,dy\,dz\, {\partial}_j \,( r_{O} {S^{j}}_i),
\label{sumcon}
\ee
which is the one used in Eq. (\ref{eom4}), which eventually leads
to Eqs. (\ref{card1}) and (\ref{card2}). Note that, with this
choice, $dF^i_O$ depends on the reference point $O$, which is why
it has a suffix $O$. On the contrary, in the second approach,
forces are not red-shifted as they are translated, and therefore
one now writes
\be
d{\cal F}^i=- dx\,dy\,dz\, {\partial}_j
{S^{j}}_i,
\ee
which is the expression used in classical theory.
The extra term that one gets, i.e. the last term on the r.h.s. of
Eq. (\ref{parder}), is now  interpreted as a contribution to the
inertia of the matter inside the cube, and is therefore shifted to
the l.h.s. of Eq. (\ref{eom3bis}), leading to Eq. (\ref{eomstr}),
and eventually to the concept of weight in Eq. (\ref{we}).

Of course, both approaches are {\it mathematically} correct.
However, as we think with the author of Ref. \cite{nord} that the
phenomenon of red-shift for forces represents a genuine physical
effect, we believe that in all situations one should accordingly
modify the sum of forces acting at different points. Therefore we
regard Eq. (\ref{sumcon}) and Eq. (\ref{ftot})
as the {\it physically correct} ones.
A further important advantage of this approach is
that it leads to a very simple concept of weight that only
involves the density of mass $\rho$ of the body, as in Eq.
(\ref{card1}). This should be contrasted with the very
complicated concept of weight in the second approach, Eq.
(\ref{we}), which explicitly involves an average of the stresses
(the latter quantity being hard to evaluate in general).

Applications of our formalism are now described in the following
sections.

\section{The case of a vessel filled with a fluid}

It is instructive to use the previous general formulae to examine
a system composed by {\it two } subsystems, i.e. a rigid vessel,
filled with  a fluid. We consider, for simplicity, the case of a
rectangular box, hanging by a thread. The box is described by an
energy-momentum tensor of the same form as in Eq. (\ref{tab}):
\be
T_{(\rm box)}^{ab}
= \rho_{(\rm box)} \, u^a u^b + S_{(\rm box)}^{ab},
\label{(3.1)}
\ee
while for the fluid one has
\be
T_{({\rm fl})}^{ab}= \rho_{({\rm
fl})} \, u^a u^b + p_{(\rm fl)}\;\delta^{ab},
\label{(3.2)}
\ee
where $p_{(\rm fl)}$ is the pressure.
If we consider the {\it total} system
formed by the box together with the fluid, the only external force
is the force $\vec{f}^{(\rm thr)}(Q)$ applied by the thread, in
the suspension point $Q$. We can determine $\vec{f}^{(\rm thr)}$
by using the First Cardinal Equation, Eq. (\ref{card1}), with
$O=Q$, and we obtain
\be
\vec{f}^{(\rm thr)}(Q)=-(M_{(\rm box)}+M_{(\rm fl)})\;\vec{{\rm g}}(Q),
\label{box1}
\label{(3.3)}
\ee
where
\be
M_{(\rm box)}\equiv \int_{(\rm box \;walls)}
\!\!\!\!\!d^3x\;\rho_{(\rm box)}\;,
\label{mbox}
\ee
and
\be
M_{(\rm fl)} \equiv  \int_{(\rm fluid)} d^3x\;\rho_{(\rm fl)}.
\label{mfl}
\ee
A comment on the above equation is now in order. Even though
the expression for $\vec{f}^{(\rm thr)}(Q)$ has {\it mathematically}
the form of the sum of two distinct contributions,
one from the box and the other from the fluid, it would be wrong to
think of the former as the weight of the {\it empty box}, i.e.
{\it without} the fluid in its interior. This is so, because when
the box is filled with the fluid, the pressure exerted by the
fluid on its internal walls causes  a small deformation of the
walls, and therefore these pressure forces make some work $W$ on
the box. This work causes a small change $\delta M_{(\rm box)}$ in
the mass of the box, of magnitude $W/c^2$, and therefore $M_{(\rm
box)} \neq M_{(\rm empty \;box)}$. However, for a very stiff box,
$\delta M_{(\rm box)}$ is extremely small and therefore, for all
practical purposes, one can identify $M_{(\rm box)}$ with $M_{(\rm
empty \;box)}$. After this identification is made, Eq.
(\ref{box1}) can be given the standard classical interpretation,
according to which the total weight of the system is the sum of
the {\it separate} weights of the box and of the fluid.

It is interesting now to repeat the analysis, by considering just
the box as the whole system. Now, in addition to $\vec{f}^{(\rm
thr)}(Q)$, the external forces acting on the box include the
pressure forces exerted by the fluid filling it. The First
Cardinal Equation, again taken for $O=Q$, now gives
\be
\vec{f}^{(\rm thr)}(Q)=-M_{(\rm box)}\;\vec{{\rm g}}(Q)-
\vec{f}^{(\rm fl)}_Q ,\label{box2} \ee where \be \vec{f}^{(\rm
fl)}_Q =\int_{\Sigma_{(\rm int)}} d^2 \sigma \;r_{Q}\; p_{(\rm
fl)} \;\hat{n}.
\label{forp}
\ee
Here, $\Sigma_{(\rm int)}$ is the
internal surface of the box, and $\hat{n}$ is the unit normal to
$\Sigma_{(\rm int)}$, pointing  inside the box, i.e. outwards with
respect to the cavity filled with fluid. By symmetry, the lateral
walls of the cavity give a vanishing net force, and therefore we
have
\be
\vec{f}^{(\rm fl)}_Q={\cal A} \;[r_{Q}(z_2)\; p_{(\rm
fl)}(z_2)- r_{Q}(z_1)\; p_{(\rm fl)}(z_1)]\;\hat{z},
\label{ffl}
\ee
where ${\cal A}$ is the area of the base of the box, while
$z_2$ and $z_1$ are the heights of the upper and lower sides of
the cavity, respectively. Note, again, that {\it the relativistic sum
of the pressures is not equal to their algebraic sum}, as it
happens in classical theory, because it involves the respective
red-shifts. Of course, Eq. (\ref{box2})  should eventually
reproduce Eq. (\ref{box1}). To see it explicitly, we note that by
the same steps leading to Eq. (\ref{eom5}), one finds that Euler's
Equations for the fluid $\nabla_a T_{({\rm fl})}^{ab}=0$ imply
\be
-\rho_{(\rm fl)} \, {\rm g}_z(Q) =  -
 \frac{d}{dz}\; ( r_{Q} p_{(\rm fl)}) .
\label{eufl}
\ee
Upon integrating the above Equation from $z_1$ to $z_2$, we find
\be
 r_{Q}(z_2)\; p_{(\rm fl)}(z_2)-
r_{Q}(z_1)\; p_{(\rm fl)}(z_1)={\rm g}_z(Q) \int_{z_1}^{z_2}
 dz\;\rho_{(\rm fl)}(z).
\label{(3.10)}
\ee
By using this result into Eq. (\ref{ffl}), we obtain
\be
\vec{f}^{(\rm fl)}_Q=\vec{{\rm g}}(Q)\, {\cal A}\,
\int_{z_1}^{z_2}
 dz\;\rho_{(\rm fl)}(z)=\vec{{\rm g}}(Q)\,M_{(\rm fl)}.
\label{ffl1}
\ee
Upon inserting this formula into Eq. (\ref{box2}), we recover Eq.
(\ref{box1}).

It is now interesting to consider the same problem from the point
of view of the alternative Eqs. (\ref{card1bis}-\ref{ftotcl}).
The final result, Eq. (\ref{box1}) will
of course be the same, but it is instructive to see how this comes about if
one considers again the problem from the point of view of the box only.
Instead of Eq. (\ref{box2}), Eqs. (\ref{card1bis}-\ref{ftotcl}) now give
\be
\vec{f}^{(\rm thr)}(Q)=-\vec{\cal P}_{(\rm box)} -
\vec{\cal{F}}^{(\rm fl)}
\label{box3}.
\ee
Here, according to Eqs. (\ref{we}) and (\ref{mij})
\be
\vec{\cal P}_{(\rm box)}
=\int_{(\rm box \;walls)}\!\!\!\!\!\! d^3 x\;
\left(\rho_{(\rm
box)}\,\delta_i^j+\frac{1}{c^2}\,S^{\;\;\;j}_{(\rm box)i}\right)
\, {\rm g}_j(z)\,\hat{x}^i ,
\label{wbox}
\ee
while, according to
Eq. (\ref{ftotcl}), for the total force exerted on the box by the
fluid we have the classical formula
\be
\vec{\cal F}^{(\rm fl)}
=\int_{\Sigma_{(\rm int)}} d^2 \sigma \;  p_{(\rm fl)}
\;\hat{n}={\cal A} \;[p_{(\rm fl)}(z_2)- p_{(\rm
fl)}(z_1)]\;\hat{z}.
\label{(3.14)}
\ee
Now, from Eq. (\ref{eufl}), we find
\begin{eqnarray}
\; & \; & p_{(\rm fl)}(z_2)-
p_{(\rm fl)}(z_1) \nonumber \\
&=& {\rm g}_z(Q) \!\!\int_{z_1}^{z_2}
\!\!\!\! dz\;\rho_{(\rm fl)} -\int_{z_1}^{z_2}\!\!\!\!dz \,p_{(\rm
fl)} \frac{d  r_{Q}}{dz} \nonumber \\
&=& {\rm
g}_z(Q)  \int_{z_1}^{z_2}\! dz \left(\rho_{(\rm fl)}
+\,\frac{1}{c^2}\;p_{(\rm fl)}\right),
\label{(3.15)}
\end{eqnarray}
where in the last passage we have used Eq. (\ref{reds}), Eq.
(\ref{acc}) and Eq. (\ref{g}) to write $d  r_{P}/dz={\rm
g}_z(Q)/c^2$. If the fluid satisfies an equation of state of the
form
\be
p_{(\rm fl)}=\gamma  \;{\rho_{(\rm fl)}}\,c^2 ,
\label{(3.16)}
\ee
we then find for $ \vec{\cal F}^{(\rm fl)}$
\be
\vec{\cal F}^{(\rm
fl)} =\vec{{\rm g}}(Q)\,(1+\gamma)\,M_{(\rm fl)} ,
\label{fclg}
\ee
a result which, at first sight, seems to contradict the
implications of Eq. (2.1). (It is interesting to note
that, if the box is filled with thermal radiation, $1+\gamma=4/3$.
This is the same ``anomalous'' factor of $4/3$ that occurred in
the classical models for the electromagnetic mass of the electron,
considered by H.A. Lorentz at the end of the nineteenth century).
Of course, there is really no contradiction, because what one
measures here is not $ \vec{\cal F}^{(\rm fl)}$ by itself, but
rather $\vec{f}^{(\rm thr)}(Q)$, which includes also the
``weight'' of the box $\vec{\cal P}_{(\rm box)}$. Now, according
to Eq. (\ref{wbox}), $\vec{\cal P}_{(\rm box)}$ can be separated
in two parts, i.e.
\be
\vec{\cal P}_{(\rm box)}=\vec{\cal P}_{(\rm
box)}^{(1)}+\vec{\cal P}_{(\rm box)}^{(2)},
\label{(3.18)}
\ee
where
\be
\vec{\cal P}_{(\rm box)}^{(1)}= \int_{(\rm box
\;walls)}\!\!\!\!\!\! d^3 x\;  \rho_{(\rm box)}\, \, \vec{\rm
g}(z),
\label{w1box}
\ee
and
\be
\vec{\cal P}_{(\rm box)}^{(2)}=
\frac{1}{c^2}\,\int_{(\rm box \;walls)}\!\!\!\!\!\! d^3 x\;
S^{\;\;\;j}_{(\rm box)i}  \, {\rm g}_j(z)\,\hat{x}^i .
\label{w2box}
\ee
Recalling the considerations following Eq.
(\ref{mfl}), for a stiff box $\vec{\cal P}_{(\rm box)}^{(1)}$ is
independent, to a high degree of precision, of the box being
filled or empty, and therefore can be interpreted as a feature of
the box, by itself. On the contrary, the second contribution
$\vec{\cal P}_{(\rm box)}^{(2)}$ depends on the stresses in the
box walls, and therefore this term is strongly affected by the
presence of the fluid, whose pressure on the inner surfaces of the
walls leads to additional stresses in the walls \footnote{In the
context of the classical model for the electron, the importance of
this contribution from the stresses of the mechanical system
needed to ensure the electron's stability, to correct the
anomalous factor of $4/3$, was first recognized by
Poincar$\acute{\rm e}$.}. We can see this  clearly by explicitly
evaluating $\vec{\cal P}_{(\rm box)}^{(2)}$. We need not perform
any extra calculations, because we can exploit the mathematical
equivalence of the two formulations of the first Cardinal Equation
to obtain $\vec{\cal P}_{(\rm box)}^{(2)}$. Upon comparing the
expression for $\vec{f}^{(\rm thr)}(Q)$ in Eq. (\ref{box2}) with
Eq. (\ref{box3}), we obtain
\be
\vec{\cal P}_{(\rm box)}^{(2)}
=(\vec{f}^{(\rm fl)}_Q-\vec{\cal{F}}^{(\rm fl)})+(M_{(\rm
box)}\;\vec{{\rm g}}(Q)-\vec{\cal P}_{(\rm box)}^{(1)}) .
\label{(3.21)}
\ee
Upon using Eqs. (\ref{ffl1}),  (\ref{fclg}), (\ref{mbox}) and
(\ref{w1box}), we then obtain
\be
\vec{\cal P}_{(\rm
box)}^{(2)}=-\gamma \,\vec{{\rm g}}(Q)\,M_{(\rm fl)}+\int_{(\rm
box \;walls)}\!\!\!\!\!\!\!\!\! d^3 x\; \rho_{(\rm box)} (\vec{\rm
g}(Q)-\vec{\rm g}(z)).
\label{(3.22)}
\ee
As we see, the first term on the r.h.s
cancels the undesirable $\gamma$-dependent contribution in Eq.
(\ref{fclg}).  However, the fact that $\vec{\cal P}_{(\rm box)}$
depends, via  this term, on the fluid, shows clearly another possible
deficiency (or peculiar property)
of Eqs. (\ref{card1bis}-\ref{we}):  for a system formed
by several bodies in contact, Eq. (\ref{card1bis}) leads to a
concept of weight for the individual bodies constituting the
system that depends strongly on the other bodies with which it
interacts. This should be contrasted with the first formulation
based on Eq. (\ref{card1}), which on the contrary permits, to a
high degree of precision (see comments following Eq. (\ref{mfl})),
to consider the weights of the individual bodies as
{\it independent} of each other.

\section{Forces on Casimir apparatuses}

We now turn to the central problem of this paper, i.e. determining
the forces that act on a Casimir apparatus suspended in the
Earth's gravitational field. For simplicity, we consider the
idealized case of a cavity consisting of two perfectly reflecting
horizontal plates, with common thickness $D$, separated by an
empty gap of width $a$. We let the coordinate system be chosen so
that the inner faces of the plates, bounding  the cavity, have
equations $z=z_1=0$ and $z=z_2=a$, respectively. Then, to leading
order ${\rm g}\, a/c^2$ (${\rm g}=|\vec{\rm g}|$), in Ref.
\cite{bimonte} we obtained the following expression for the
nonvanishing components of the Casimir energy-momentum tensor
$\langle T^{ab}_{\rm (C)}\rangle$ in the local coordinate system
of Eq. (\ref{un}):
\begin{widetext}\begin{eqnarray}
\langle T^{00}_{\rm (C)}\rangle(z) &=& -\frac{\pi^2 \hbar}{c\,a^4}
\left[\frac{1}{720}+\frac{2\, {\rm
g}\,a}{c^2}\left(\frac{1}{1200}-\frac{1}{3600}
\frac{z}{a}-\frac{\cot(\pi z/a) \csc^2(\pi z/a)}{240 \pi }\right)
\right], \label{T00}\\
\langle T^{11}_{\rm (C)}\rangle(z)&=&\langle T^{22}_{\rm
(C)}\rangle (z)=\frac{\pi^2 \hbar
c}{a^4}\left[\frac{1}{720}+\frac{2\, {\rm
g}\,a}{c^2}\left(\frac{1}{3600}-\frac{1}{1800}\frac{z}{a}
-\frac{\cot(\pi z/a) \csc^2(\pi z/a)}{120 \pi}
\right)\right],  \\
\langle T^{33}_{\rm (C)}\rangle (z) &=& -\frac{\pi^2 \hbar
c}{a^4}\left[\frac{1}{240}+\frac{2\, {\rm
g}\,a}{c^2}\frac{1}{720}\left(1-2\frac{z}{a}\right)\right].
\label{T33}
\end{eqnarray}
\end{widetext}
These equations, obtained after performing a very difficult
calculation, based upon the covariant geodesic point separation
\cite{Chri76, Chri78}, are on firm ground because the following
consistency checks are satisfied: \vskip 0.3cm \noindent (i) The
photon and ghost Green functions used to build the Casimir
energy-momentum tensor obey the perfect-conductor boundary
conditions. \vskip 0.3cm \noindent (ii) The Ward identity $G_{\mu
\nu'}^{\; \; \; \; \; ;\mu}+G_{;\nu'}=0$, relating covariant
derivatives of photon and ghost Green functions, has been checked
to first order in ${\rm g}a/c^{2}$. \vskip 0.3cm \noindent (iii)
The Casimir energy-momentum tensor satisfies, to first order in
${\rm g}a/c^{2}$, the covariant conservation law \be \nabla_a
\langle T^{ab}_{\rm (C)} \rangle=0. \label{nabflu} \ee Had we
considered instead the case of a strong gravitational field, it
would have been impossible to perform a Fourier analysis of Green
functions as in Ref. \cite{bimonte}. The resulting evaluation of
the  energy-momentum tensor remains an open problem.

We note that, since $\langle T^{0i}_{\rm (C)} \rangle (z) =
\langle T^{i0}_{\rm (C)} \rangle(z)=0$, the Casimir
energy-momentum tensor has the form corresponding to a ``body'' at
rest, as in Eq. (\ref{tab}). Therefore, all theorems derived in
Sec. II automatically apply to a Casimir apparatus. In particular,
Eq. (\ref{mass}) holds, and therefore the Casimir apparatus has a
passive gravitational mass $M_{{\rm(C)}}$ which is equal to its
total inertia: \be M_{{\rm (C)}}= \int_{\rm cavity}
d^3x\;\rho_{\rm(C)}. \label{(4.5)} \ee Recalling that, according
to Eq. (\ref{tab}) \be \rho_{\rm(C)}= \frac{1}{c^4}\,\langle
T^{ab}_{\rm (C)} \rangle\,u_a u_b , \label{(4.6)} \ee we obtain
from Eq. (\ref{T00}) \be M_{\rm(C)}= -\frac{\pi^2 \hbar}{720\,c\,
a^3}\,{\cal A}+O\,({\rm g} a/c^2)=\frac{E_{\rm(C)}}{c^2}+O\,({\rm
g} a/c^2), \label{rhoc} \ee in agreement with the findings for
$M_{\rm (C)}$ in Ref. \cite{Full07}.

\subsection{A suspended rigid Casimir cavity}

In the first setup we consider, the plates are rigidly connected
to each other, forming a unique rigid system, supported by a
thread. By steps similar to those used in Sec. II, we obtain an
Equation analogous to Eq. (\ref{box1}) for the force
$\vec{f}^{(\rm thr)}(Q)$ required to support the cavity:
\be
\vec{f}^{(\rm thr)}(Q)=-M_{(\rm box)}\vec{{\rm g}}(Q)
-M_{\rm(C)}\;\vec{{\rm g}}(Q),
\label{cas0}
\ee
where $M_{(\rm box)}$ is defined as in Eq. (\ref{mbox}). After bearing
in mind the observations made after Eq. (\ref{mfl}) on the
interpretation of $M_{(\rm box)}$, one can think of the second
term on the r.h.s. of Eq. (\ref{cas0}) as the weight
$\vec{P}^{\rm(C)}$ of the ``Casimir mass''. Using Eq.
(\ref{rhoc}), we obtain to leading order in ${\rm g} a/c^2$
\be
\vec{P}^{\rm(C)}\approx- \frac{\pi^{2}}{720} \frac{{\cal
A}{\hbar}}{c a^{3}} \vec{{\rm g}}
=\frac{E_{\rm(C)}}{c^2}\,\vec{{\rm g}},
\label{wepc}
\ee
in agreement with the weak Equivalence Principle. It is interesting
to see how the same result is obtained, if we consider the forces
acting only on the rigid walls bounding  the cavity, analogously
to what was done in Sec. II, for the case of a rigid box filled
with a fluid. Again, following the same steps that led to Eqs.
(\ref{box2}) and (\ref{forp}), we obtain
\be
\vec{f}^{(\rm thr)}(Q)
=-M_{(\rm box)}\;\vec{{\rm g}}(Q)-\int_{\Sigma_{(\rm
int)}}\!\!\!\! d^2 \sigma \;r_{Q}\;\langle T^{ij}_{\rm (C)}
\rangle \;\hat{n}_j\; \hat{x}_i .
\label{cas2}
\ee
Note again the
presence of the red-shift $r_Q$ multiplying the Casimir stresses
in the integral on the r.h.s., which is crucial to obtain the
right answer, as we shall now see. By symmetry, the lateral walls
of the cavity give a net vanishing contribution to the integral on
the r.h.s. of the above Equation, and therefore we have
\begin{eqnarray}
\; & \; & \int_{\Sigma_{(\rm int)}}\!\!\!\! d^2 \sigma \;r_{Q}\;\langle
T^{ij}_{\rm (C)} \rangle\;\hat{n}_{j} \hat{x}_{i}
={\cal A}\Bigr[\,r_Q(z_2)\,\langle T^{33}_{\rm (C)}\rangle(z_2)
\nonumber \\
&-& r_Q(z_1)\,\langle T^{33}_{\rm (C)}\rangle(z_1)\Bigr]\hat{z}.
\label{pcasr}
\end{eqnarray}
Now we see from Eq. (\ref{reds}) that, to leading
order in ${\rm g} z/c^2$, the red-shift $r_{Q}$ is
\be
r_Q(z) \approx 1+ \frac{{\rm g}}{c^2}(z-z_Q).
\label{appred}
\ee
By using this formula in Eq.
(\ref{pcasr}), together with the expression for $\langle
T^{33}_{\rm (C)}\rangle$ in Eq. (\ref{T33}), we find that the
quantity between square brackets in Eq. (\ref{pcasr}) is equal to
\be
-\frac{\pi^2 \hbar c}{a^4}\left[\frac{{\rm
g}}{240\,c^2}(z_2-z_1)-\frac{4 \,{\rm
g}}{720\,c^2}(z_2-z_1)\right]=\frac{\pi^2 \hbar c}{720\,a^3}{\rm g}.
\label{(4.13)}
\ee
Upon using this expression into Eq. (\ref{cas2}), we
recover the same result as Eq. (\ref{wepc}).

\subsection{Disconnected plates with separate mounts}

In the second setup that we wish to consider, the two plates are
{\it disconnected}, and are supported by two separate mounts. If
the mounts are connected to the outer faces of plates, the forces
$\vec{f}_1$ and $\vec{f}_2$ that support the plates are applied at
points $Q_1$ and $Q_2$, with heights $w_2=a+D$ and $w_1=-D$,
respectively. It should be remarked that the forces $\vec{f}_1$
and $\vec{f}_2$ can only be determined by
using the explicit expression of $\langle T^{33}_{(\rm C)}\rangle$
in Eq. (\ref{T33}). Upon applying the first Cardinal Equation to
each plate separately, we find for the force $\vec{f}_{I}$ ($I=1,2$)
\be
\vec{f}_I=-\vec{P}^{(I)}_{Q_I}
-\vec{f}_{Q_I}^{\rm(C)},
\label{cas1}
\ee
where $\vec{P}^{(I)}_{Q_I}$ is the weight of the $I$-th plate, i.e.
\be
\vec{P}^{(I)}_{Q_I}=\vec{{\rm g}}(w_I)\,\int_{A_I} d^3x\,\rho_I \,
\label{(4.15)}
\ee
while $\vec{f}_I^{\rm(C)}$ is the contribution from the
Casimir pressure
\be
\vec{f}_{Q_I}^{\rm(C)}\,=\,\int\!\! dx\,dy
\;r_{Q_I}(z_I)\;\langle T^{ij}_{\rm (C)}\rangle(z_I)
\;\hat{n}_j\;\hat{x}_i .
\label{(4.16)}
\ee
On using Eq. (\ref{appred}) and the
expression for $\langle T^{33}_{\rm (C)}\rangle(z)$ in Eq.
(\ref{T33}), we obtain for the upper plate, to order ${\rm g}/c^2$:
\be
\vec{f}_{Q_2}^{\rm(C)}\,\approx\,- \frac{\pi^2}{240}
\frac{{\cal A}{\hbar} c}{a^{4}} \,\left[1-\frac{\rm g}{c^2}
\left(D+\frac{2}{3}\,a \right)\right]\,\hat{z},
\label{(4.17)}
\ee
while for the lower plate we get
\be
\vec{f}_{Q_1}^{\rm(C)}\,\approx\,\frac{\pi^2}{240} \frac{{\cal
A}{\hbar}c}{a^{4}} \left[1+\frac{\rm g}{c^2}\left(D+\frac{2}{3}\,a
\right)\right]\,\hat{z}.
\label{(4.18)}
\ee
We note that both forces depend on the thickness of the plates. It is
useful to remark also that the previous result for the total force measured
by a rigid cavity, Eq. (\ref{wepc}), is recovered if the forces
${\vec f}_{Q_{1}}^{(C)}$ and ${\vec f}_{Q_{2}}^{(C)}$ are added,
say at $Q_{2}$, using the relativistic law in Sec. II, because then
\begin{eqnarray}
&&\vec{f}_{Q_2}^{\rm(C)}+r_{Q_2}(Q_1)\vec{f}_{Q_1}^{\rm(C)}\approx
F_{(\rm C)}\left\{-\left[1-\frac{\rm g}{c^2}\left(D+\frac{2}{3}\,a
\right)\right]+\right.\nonumber\\
&&+\left.\left[1-\frac{\rm g}{c^2}\left(2\,D+\,a
\right)\right]\left[1+\frac{\rm g}{c^2}\left(D+\frac{2}{3}\,a
\right)\right]\right\}\,\hat{z}\approx\nonumber\\
&&\approx \frac{1}{3}\frac{\rm g}{c^2}\,F_{(\rm
C)}\;\hat{z}=\frac{E_{\rm(C)}}{c^2}\,\vec{{\rm g}},
\label{(4.19)}
\end{eqnarray}
which is the same result as Eq. (\ref{wepc}).

\section{Concluding remarks}

Einstein's Equivalence Principle, according to which the Laws of
Physics in a uniform gravitational field are the same as in a
uniformly accelerated frame, is one of the most powerful and
general principles of Physics. Initially formulated within the
context of Classical Physics, it is currently regarded as a
universal principle, which retains its validity also in the realm
of Quantum Physics. Among the quantum phenomena, one of the most
fundamental is that of vacuum fluctuations, with the associated
unavoidable content of energy. It would clearly be of great
importance to test by experiments whether this quantum vacuum
energy conforms to the Equivalence Principle. A convincing way  to
do that would be to verify whether the energy of vacuum
fluctuations existing in a cavity with reflecting walls, i.e. the
Casimir energy $E_{(\rm C)}$, gravitates as other conventional
forms of matter-energy. While the feasibility of such an
experiment by  current weak-force measurement devices was
discussed in Ref. \cite{cal}, the present paper has analyzed in
detail, from the point of view of General Relativity, the
mechanical forces in a Casimir apparatus suspended in the Earth's
gravitational field. This is an essential step, because these
forces are the quantities to be confronted with real experiments.
For that purpose, we have derived a set of Cardinal Equations
giving the conditions for mechanical equilibrium for any extended
body, satisfying the covariant balance of energy and momentum
between body and external fields, at rest in a uniform
gravitational field. The key feature of these equations is that,
in a gravitational field, forces are subject to red-shifts, a
phenomenon originally discovered by Nordtvedt \cite{nord}, using
heuristic arguments based on the Equivalence Principle.
Consideration of this phenomenon is essential in order to obtain
the correct values for the forces occurring in an intrinsically
relativistic system, such as a Casimir apparatus. On the basis of
these Cardinal Equations, we proved rigorously that, for the case
of a rigid cavity, the weight associated with the Casimir energy
$E_{(\rm C)}$ is equal to ${\vec {\rm g}}\,E_{(\rm C)}/c^2$, as
expected. Moreover, we considered the case of a Casimir cavity
consisting of two disconnected plates, supported by separate
mounts. Also for this case the general Cardinal Equations provide
the relativistically correct expressions for the forces exerted by
the mounts on the plates.

Encouraging agreement is also found between
our force formulae, relying upon energy-momentum methods, and the
force formulae in Ref. \cite{Full07}, which rely instead upon
variational methods pioneered by Schwinger \cite{Schw51, Schw53}.

\acknowledgments We are grateful
to the authors of Ref. \cite{Full07} for having sent us
their work, prior to publication. This has substantially motivated
the research described in our paper. The work of L. Rosa has been
partially supported by PRIN {\it FISICA ASTROPARTICELLARE}.

\end{document}